\documentclass[%
 jmp,
 pre,%
 amsmath,amssymb,
 preprint,%
author-year,%
author-numerical,%
superscriptaddress,
]{revtex4-1}
\usepackage[dvipdfmx]{graphicx}
\usepackage{dcolumn}
\usepackage{bm}
\usepackage[dvips]{color}

\usepackage{algorithm}
\usepackage{algpseudocode}
\usepackage[dvipdfmx,
setpagesize=false,
bookmarks=true, 
bookmarksnumbered=true,
bookmarkstype=toc=true, 
colorlinks=true, 
urlcolor=blue,
linkcolor=blue
]{hyperref}

\def\vc(#1){\mbox{\boldmath $#1$}}

\def\p{\partial}
\def\vOmega{\mbox{\boldmath $\Omega$}}
\def\vtheta{\mbox{\boldmath $\theta$}}
\def\vTheta{{\bf \Theta}}

\def\vPsi{{\bf \Psi}}
\def\vlambda{\mbox{\boldmath $\lambda$}}
\def\vzero{\mbox{\boldmath $0$}}
\def\vD{\mbox{\boldmath $D$}}
\def\vx{\mbox{\boldmath $x$}}
\def\vX{\mbox{\boldmath $X$}}

\def\vf{\mbox{\boldmath $f$}}
\def\vF{\mbox{\boldmath $F$}}
\def\vh{\mbox{\boldmath $h$}}
\def\vH{\mbox{\boldmath $H$}}
\def\vxi{\mbox{\boldmath $\xi$}}
\def\vzeta{\mbox{\boldmath $\zeta$}}
\def\vgamma{\mbox{\boldmath $\gamma$}}

\def\vr{\mbox{\boldmath $r$}}

\def\vq{\mbox{\boldmath $q$}}

\def\vz{\mbox{\boldmath $z$}}
\def\vA{\mbox{\boldmath $A$}}
\def\va{\mbox{\boldmath $a$}}

\def\prob{p}
\def\hesse{{\boldsymbol H}}

\def\rI{I}
\def\rII{I\hspace{-.1em}I}
\def\rIII{I\hspace{-.1em}I\hspace{-.1em}I}

\newcommand{\EquationHelper}[2]{ 
\begin{equation}
#1
\label{#2} 
\end{equation} 
}

\begin{document}

\title[Data assimilation for massive autonomous systems based on second-order adjoint method]{Data assimilation for massive autonomous systems based on second-order adjoint method}
\author{Shin-ichi Ito}
\affiliation{Earthquake Research Institute, The University of Tokyo, Japan}
\author{Hiromichi Nagao}
\affiliation{Earthquake Research Institute, The University of Tokyo, Japan}
\affiliation{Graduate School of Information Science and Technology, The University of Tokyo, Japan}
\author{Akinori Yamanaka}
\affiliation{Division of Advanced Mechanical Systems Engineering, Institute of Engineering, Tokyo University of Agriculture and Technology, Japan}
\author{Yuhki Tsukada}
\affiliation{Graduate School of Engineering, Nagoya University, Japan}
\author{Toshiyuki Koyama}
\affiliation{Graduate School of Engineering, Nagoya University, Japan}
\author{Masayuki Kano}
\affiliation{Earthquake Research Institute,The University of Tokyo, Japan}
\author{Junya Inoue}
\affiliation{School of Engineering, The University of Tokyo, Japan}

\date{\today}
\begin{abstract}
Data assimilation (DA) is a fundamental computational technique that integrates numerical simulation models and observation data on the basis of Bayesian statistics. Originally developed for meteorology, especially weather forecasting, DA is now an accepted technique in various scientific fields. One key issue that remains controversial is the implementation of DA in massive simulation models under limited computation time and resources. In this paper, we propose an adjoint-based DA method for massive autonomous models that produces optimum estimates and their uncertainties within practical computation time and resource constraints. The uncertainties are given as several diagonal components of an inverse Hessian matrix, which is the covariance matrix of a normal distribution that approximates the target posterior probability density function in the neighborhood of the optimum. Conventional algorithms for deriving the inverse Hessian matrix require $O(CN^2+N^3)$ computations and $O(N^2)$ memory, where $N$ is the number of degrees of freedom of a given autonomous system and $C$ is the number of computations needed to simulate time series of suitable length. The proposed method using a second-order adjoint method allows us to directly evaluate the diagonal components of the inverse Hessian matrix without computing all of its components. This drastically reduces the number of computations to $O(C)$ and the amount of memory to $O(N)$ for each diagonal component. The proposed method is validated through numerical tests using a massive two-dimensional Kobayashi's phase-field model. We confirm that the proposed method correctly reproduces the parameter and initial state assumed in advance, and successfully evaluates the uncertainty of the parameter. Such information regarding uncertainty is valuable, as it can be used to optimize the design of experiments.
\end{abstract}

\pacs{aaaaa}


\maketitle
\section{Introduction}

Determining the model parameters and initial states of simulation models is an important task in various scientific fields, as it enables the temporal evolution of the target system to be observed. However, in many practical cases, this procedure is somewhat complex, because it is often impossible to observe the parameters and initial states experimentally. In materials engineering, for example, phase-field (PF) models are often used to simulate the evolution of microstructures during the processes of solidification and phase transformation~\cite{ kobayashi1993modeling, Boettinger2002PHASE, Chen2002PHASE, Tsukada2008Phase, Shimokawabe2011Peta,Ohno2013Existence, Takaki2014Phase}. PF models phenomenologically describe the dynamics of phases using field variables that evolve in time depending on the gradient of the total free energy. Since a PF model usually requires a huge number of grid points to discretize the field variables, the computational cost tends to be prohibitive. Nonetheless, PF models are accepted beyond the field of materials engineering, such as in hydrodynamics~\cite{Jacqmin1999Calculation, DeMenech2006Modeling, Lee2011OntheLong}, as they can be employed to model phases and their dynamics using mathematical expressions that are easy to manipulate. In PF models, the parameters and initial states perfectly determine the dynamical properties of the phases. However, various limitations in practical experiments sometimes prevent such parameters and initial states from being estimated.

Data assimilation (DA) is a computational technique that integrates numerical simulation models and observational data on the basis of Bayesian statistics. Thus, DA enables the parameters and initial states of PF models to be estimated by systematically extracting as much information as possible from the given observational/experimental data. The process of DA evaluates a probability density function (PDF) (or the ``posterior PDF,'' to be precise) of the unknown parameters and unobservable states that is conditional on the given observation data~\cite{reich2015probabilistic}. DA was originally developed in the fields of meteorology and oceanography~\cite{kalnay2003atmospheric, Tuyuki2007Recent, Ghil1991Data}, but is now applied in areas such as seismology, marketing science, and industrial science~\cite{Kano2015Real, Maeda2015Successive, Motohashi2012or, Sasaki2016prep}. Several sequential Bayesian filters and other non-sequential estimation methods have been used in DA. Common sequential Bayesian filters such as the ensemble Kalman filter~\cite{evensen2003ensemble, houtekamer1998data, Ueno2007Application} and particle filter~\cite{kitagawa2010introduction, doucet2000sequential, Nagao2013TimeSeries} estimate the target posterior PDF using Bayes' theorem. This approximation is formed using an ensemble of realizations, meaning that the computational cost is proportional to the number of realizations. The implementation of one sequential Bayesian filter on a given simulation model is not especially complex, and a sufficiently accurate estimate of the posterior PDF can be achieved when the number of degrees of freedom $N$ of the simulation model is sufficiently small. However, Bayesian filters become inefficient when applied to massive simulation models, as the number of realizations required to obtain a converged posterior PDF is proportional to $e^{O(N)}$. Unlike sequential Bayesian filters, adjoint methods~\cite{Lewis1985TheUse, LeDimet1986Variational, iri1991FAD} directly determine the optimum solution using a gradient method to maximize the target posterior PDF. Although this achieves a drastic reduction in the computational cost, the ordinary adjoint method cannot evaluate the uncertainty in its estimations, something which sequential Bayesian filters obtain in a straightforward manner. Such uncertainties provide valuable information related to both the estimations and the optimum solution. For example, the uncertainties provide feedback for the experimental design that helps to identify the parameters of interest with the required accuracy. The quantification of uncertainty is currently a very important issue in the application of DA to massive simulation models.

This paper describes an adjoint-based DA methodology that simultaneously estimates the optimum solution and its uncertainty, and is capable of being applied to simulation models with a huge number of degrees of freedom. We first construct a method to estimate the parameters and initial states involved in an autonomous system, and then validate our approach using a PF model as a testbed. Section~\ref{chap2} introduces the formulation of an adjoint-based DA method to simultaneously obtain an estimation and its uncertainty using second-order information of the posterior PDF. Section~\ref{chap3} describes the formulation of an estimation test using synthetic data generated from the time series given by Kobayashi's PF model~\cite{kobayashi1993modeling}. Section~\ref{chap4} presents and discusses the results of estimation tests, and Section~\ref{chap5} concludes this paper by summarizing the results of this study.

\section{Method\label{chap2}}
\subsection{State-space model and cost function\label{chap2A}}
DA based on Bayesian statistics always starts by defining a state-space model, which consists of a system model and an observation model. The system model describes how a state vector evolves over time in accordance with a given simulation model. The state vector contains all time-dependent variables used in the simulation model and sometimes the model parameters.

Suppose an autonomous simulation model is given by $\p\vz\slash\p t= \vA(\vz; \va)$, where $\vz(t)\in\mathbb{R}^{N_{z}}$ denotes a time-dependent variable and $\vA: \mathbb{R}^{N_{z}}\rightarrow\mathbb{R}^{N_{z}}$ is a function of $\vz$ and a time-invariant parameter vector $\va\in\mathbb{R}^{N_{a}}$. The system model describes the time evolution of a state vector consisting of $\vz$ and $\va$. Let $\vX(t)=\left(\vz^{\top}, \va^{\top}\right)^{\top} \in \mathbb{R}^{N}$ be the state vector, where $\bullet^{\top}$ denotes the transpose of $\bullet$ and $N=N_{z}+N_{a}$. Since $\va$ is time-invariant, i.e., $\p\va\slash\p t= \vzero$, the system model can be represented by
\EquationHelper{
\frac{\p \vX}{\partial t} = \vf\left(\vX\right),
}{sect2.1.1}
where $\vf : \mathbb{R}^{N}\rightarrow\mathbb{R}^{N}$ is defined as $f_{i}=A_{i}$ for $1\le i \le N_{z}$ and $f_{i}=0$ for $N_{z} + 1\le i \le N$.

The observation model describes how $\vX(t)$ relates to a time series of observation data $\vD(t)\in \mathbb{R}^{K}$, where $K$ denotes the dimension of the observations. Considering that the data include noise, the observation model can be described as  
\EquationHelper{
\vD = \check{\vh} \left( \vX\right) + \vOmega, 
}{sect2.1.2}
where $\check{\vh} : \mathbb{R}^{N}\rightarrow \mathbb{R}^{K}$ is an observation operator that outputs quantities from $\vX$ comparable with the data and $\vOmega(t)$ denotes observation noise. In this paper, we assume that $\vf$ and $\check{\vh}$ are nonlinear functions, and $\vOmega$ is white noise that follows a normal distribution with a diagonal covariance matrix. Our purpose is to obtain the optimum initial state $\vX(0)$ together with the uncertainties of the variables of interest.

In consideration of PF models, we also assume that $\vX(0)$ is constrained by  
\EquationHelper{
X_{i}^{\text{Lower}} < X_{i}(0) < X_{i}^{\text{Upper}} \quad (i=1,\cdots,N),
}{sect2.1.3}
where $\vX^{\text{Lower}} \in \mathbb{R}^{N}$ and $\vX^{\text{Upper}} \in \mathbb{R}^{N}$ denote the lower and upper bounds of $\vX(0)$, respectively.

To simplify our formulation, we normalize $\vX$ as  
\EquationHelper{
\theta_{i}(t) = \frac{ X_{i}(t)-X_{i}^{\text{Lower}} }{ X_{i}^{\text{Upper}}-X_{i}^{\text{Lower}} } \quad (i=1,\cdots,N).
}{normalized}
This leads to the following $\vtheta$-dependent forms of Eqs.~\eqref{sect2.1.1}-\eqref{sect2.1.3}:  
\EquationHelper{
\frac{\p \vtheta}{\partial t} = \vF(\vtheta),
}{sect2.1.4}
\EquationHelper{
\vD = \vh \left( \vtheta \right) + \vOmega, 
}{sect2.1.5}
\EquationHelper{
0 < \Theta_{i} < 1 \quad (i=1,\cdots,N),
}{sect2.1.6}
where $F_{i}(\vtheta)=f_{i}(\vX)\slash\left( X_{i}^{\text{Upper}}-X_{i}^{\text{Lower}} \right)$, $\vh\left(\vtheta\right)$ is an observation operator after transforming $\vX$ to $\vtheta$, and $\vTheta = \vtheta(0)$.

Bayes' theorem states that a conditional PDF $\prob\left(\vTheta|\vD\right)$, which is called the posterior PDF, can be described as  
\EquationHelper{
\prob(\vTheta|\vD) = \frac{\prob(\vTheta)\prob\left( \vD|\vTheta\right)}{\prob(\vD)},
}{sect2.1.7}
where $\prob(\vTheta)$ and $\prob\left( \vD|\vTheta\right)$ are called the prior PDF and likelihood, respectively. Note that $\prob(\vD)$ is constant, since $\vD$ is a definite vector. Thus, Eq.~\eqref{sect2.1.7} implies that $\prob\left(\vTheta|\vD\right)$ is proportional to a product of the prior PDF and the likelihood.

The prior PDF contains prior information provided by experience and intuition. If we suppose that this prior information is the constraint condition given by Eq.~\eqref{sect2.1.6} and that $\Theta_{i}$ is independent for each $i$, $\prob(\vTheta)$ is given by a product of prior PDFs of $\Theta_{i}$,  
\EquationHelper{
\prob(\vTheta) = \prod_{i=1}^{N}\prob(\Theta_{i}),
}{prior} where
\EquationHelper{
\prob(\Theta_{i}) = \left\{
\begin{array}{ccc}
1 & \text{for} & 0 < \Theta_{i} < 1 \\
0 & \; & \text{otherwise}.
\end{array}\right.
}{sect2.1.8}

When observation data are obtained at $t=t_{1}, t_{2}, \cdots, t_{n}$, $\prob\left( \vD|\vTheta\right)$ can be written as  
\EquationHelper{
\prob\left( \vD|\vTheta\right) = \prod_{s=1}^{n}\prod_{k=1}^{K}\prob\left(\Omega_{k}(t_{s})\right),
}{likelihood}
where 
\EquationHelper{
\prob\left(\Omega_{k}(t_{s})\right) = \frac{1}{\sqrt{2\pi\sigma_{k}^2}}\exp\left[ -\frac{
\left\{  D_{k}(t_{s}) - h_{k}\left( \vtheta(t_{s})\right) \right\}^2
}{2\sigma_{k}^{2}} \right],
}{sect2.1.9}
and $\sigma_{k}$ is the standard deviation of $\Omega_{k}$ ($k=1,\cdots,K$). We consider $\sigma_{k}$ to be a hyper-parameter.

This paper defines the optimum solution $\hat{\vTheta}$ to be the $\vTheta$ that maximizes the posterior PDF $\prob\left(\vTheta|\vD\right)$. For the convenience of numerical computation, we aim to minimize a cost function  
\EquationHelper{
\begin{aligned}
J = \sum_{s=1}^{n}\sum_{k=1}^{K}\left[ \frac{\log(2\pi\sigma_{k}^{2})}{2} 
+  \frac{ \left\{  D_{k}(t_{s})  - h_{k}\left( \vtheta(t_{s})\right)\right\}^2 } {2\sigma^{2}_{k}}\right] \text{  subject to  } 0 < \Theta_{i} < 1,
\end{aligned}
}{sect2.1.10} which comes from a negative logarithmic posterior PDF, i.e., $\prob(\vTheta|\vD) \propto e^{-J}$, to find $\hat{\vTheta}$, rather than maximizing $\prob\left(\vTheta|\vD\right)$.
The constraint in Eq.~\eqref{sect2.1.10} arises from the term $-\log \prob\left(\vTheta\right)$, which appears when calculating $-\log \prob(\vTheta|\vD)$.

An optimum solution $\hat{\sigma}_{k}$ for $\sigma_{k}$ can be determined as follows. By letting $\p J \slash \p \sigma_{k} = 0$, we obtain  
\EquationHelper{
\sigma_{k} = \sqrt{ \frac{1}{n}\sum_{s=1}^{n}\left[ D_{k}(t_{s}) - h_{k}\left( \vtheta(t_{s})\right)   \right]^2 }.
}{sect2.1.11}
Then, $\hat{\sigma}_{k}$ is obtained by substituting $\vtheta$, which is calculated by Eq.~\eqref{sect2.1.4} using $\vtheta(0)=\hat{\vTheta}$, into Eq.~\eqref{sect2.1.11}.

\subsection{Optimization via an adjoint method\label{chap2B}}
Typically, $J$ is optimized using a gradient method such as steepest gradient descent, the nonlinear conjugate gradient method, or the Limited-memory Broyden-Fletcher-Goldfarb-Shanno (LBFGS) method~\cite{nocedal1980updating}. Gradient methods require $\p J \slash \p \vTheta$ to update $J$, but it is difficult to calculate this quantity because $J$ does not explicitly include $\vTheta$, as seen in Eq. (13). Generally, $J$ is fully determined by setting $\vTheta=\vtheta(0)$ through Eq.~\eqref{sect2.1.4}, so that $J$ must be written as a function of $\vTheta$, i.e., $J(\vTheta)$. According to Eq.~\eqref{sect2.1.10}, $J$ is also a function of $\vtheta\left(t_{s}\right)$ $(s=1,\cdots,n)$ that satisfies Eq.~\eqref{sect2.1.4}, not including $\vTheta$. Summarizing these two expressions for $J$:  
\EquationHelper{
J(\vTheta) = \int_{0}^{t_{f}} dt\; \mathcal{J}(\vtheta) + \int_{0}^{t_{f}} dt\; \vlambda^{\top}\left(\vF - \frac{\p\vtheta}{\p t}\right),
}{sect2.2.1}
where $t_{f}$ is an arbitrary time later than $t_{n}$, and $\vlambda(t)\in \mathbb{R}^{N}$ denotes a vector of the Lagrange multipliers that impose Eq.\eqref{sect2.1.4} as the constraint condition of $\vtheta$. $\mathcal{J}$ is the time-dependent function  
\EquationHelper{
\mathcal{J}(\vtheta) = 
\sum_{s=1}^{n} \delta (t-t_s)\sum_{k=1}^{K}\left[ \frac{\log(2\pi\sigma_{k}^{2})}{2} 
+  \frac{ \left\{  D_{k}(t)  - h_{k}\left( \vtheta(t)\right)\right\}^2 } {2\sigma^{2}_{k}}\right],
}{sect2.2.2}
that satisfies $J = \int_{0}^{t_f} dt\;\mathcal{J}$, where $\delta(t)$ denotes the Dirac delta function. Taking a variation of Eq.~\eqref{sect2.2.1}, we have a time evolution equation for $\vlambda$:  
\EquationHelper{
\frac{\p\vlambda}{\p t} + \left(\frac{\p \vF}{\p \vtheta}\right)^{\top}\vlambda + \frac{\p\mathcal{J}}{\p\vtheta} = \vzero,
}{sect2.2.3}
where
\EquationHelper{
\vlambda(0) = \frac{\p J}{\p \vTheta},
}{sect2.2.4}
\EquationHelper{
\vlambda(t_{f}) = \vzero.
}{sect2.2.4-1}
Details of the derivation can be found in~\cite{LeDimet1986Variational,Wang1992Second}. Solving Eq.~\eqref{sect2.2.3} backwardly in time with the condition in Eq.~\eqref{sect2.2.4-1}, we obtain the objective $\p J \slash \p \vTheta$ as $\vlambda(0)$. Such a procedure to obtain the gradient of the cost function using the adjoint equation (Eq.~\eqref{sect2.2.3}) is called the adjoint method.

When we apply the adjoint method to our problem, a variable transformation is needed in the process of updating $\vTheta$ based on a gradient method, since $\vTheta$ has the constraint shown in Eq.~\eqref{sect2.1.6}. The variable transformation  
\EquationHelper{
\Psi_{i} = \log \Theta_{i} -\log\left(1-\Theta_{i} \right) \quad (i=1,\cdots,N)
}{variabletrainsformation}
converts the constrained optimization problem of $\vTheta$ into an unconstrained one with respect to $\vPsi$. The update procedure is as follows. After obtaining $\p J \slash \p \vTheta$ by the adjoint method based on Eqs.~\eqref{sect2.2.3}-\eqref{sect2.2.4-1}, we convert $\vTheta$ to $\vPsi$ using Eq.~\eqref{variabletrainsformation} and $\p J \slash \p \vTheta$ to $\p J \slash \p \vPsi$ as  
\EquationHelper{
\frac{\p J}{\p\Psi_{i}} = \Theta_{i}\left(1-\Theta_{i}\right) \frac{\p J}{\p\Theta_{i}} \quad (i=1,\cdots,N).
}{gradpsi}
Using this formulation to update $\vPsi$, we can obtain an updated $\vTheta$ from the inverse transformation of Eq.~\eqref{variabletrainsformation}:  
\EquationHelper{
\Theta_{i} = \frac{1}{1+\exp\left(-\Psi_{i}\right)} \quad (i=1,\cdots,N).
}{invvariabletrainsformation}
Although this update procedure does not allow $\Theta_{i}$ to be exactly $0$ or $1$, owing to the definition of Eq.~\eqref{invvariabletrainsformation}, $\Theta_{i}$ can be sufficiently close to $0$ or $1$ to pose no problem in practical cases.

The adjoint method calculates $\p J \slash \p \vTheta$ for a fixed $\sigma_{k}$, so that an optimization of $\sigma_{k}$ is to be done at the same time as $\vTheta$ by substituting Eq.~\eqref{sect2.1.11} into Eq.~\eqref{sect2.2.2} every time $\vTheta$ is updated.

The advantages of the adjoint method over sequential Bayesian filters are that only $O(C)$ computations and $O(N)$ memory are required to find $\hat{\vTheta}$, where $C$ is the number of computations needed to run the given simulation model from $t=0$ to $t=t_{f}$.

\subsection{Evaluation of uncertainty via a second-order adjoint method\label{chap2D}}
The adjoint method described in Section~\ref{chap2B} gives the optimum solution $\hat{\vTheta}$ that maximizes $\prob\left(\vTheta|\vD\right)$. However, it does not provide information about the behavior of $\prob\left(\vTheta|\vD\right)$ in the neighborhood of $\vTheta=\hat{\vTheta}$, which reflects the uncertainty in the estimation of $\hat{\vTheta}$. To extract such information, another procedure must be implemented on the adjoint method. Considering that $\p J \slash \p\vTheta |_{\vTheta=\hat{\vTheta}}=\vzero$, the Taylor expansion of $J$ with respect to $\vTheta-\hat{\vTheta}$ is  
\EquationHelper{
J(\vTheta) \sim J(\hat{\vTheta}) + \frac{1}{2} (\vTheta-\hat{\vTheta})^{\top} \hesse (\vTheta-\hat{\vTheta} ),
}{sect2.2.5}
where terms of order higher than three have been neglected, and $\hesse$ is a Hessian matrix given by  
\EquationHelper{
H_{i,j} =  \left.\frac{\p^{2} J}{\p\Theta_{i}\p\Theta_{j}}\right|_{\vTheta=\hat{\vTheta}} \quad(i,j=1,\cdots,N).
}{sect2.2.6}
We normalize $\prob(\vTheta|\vD)\propto e^{-J}$ into which Eq.~\eqref{sect2.2.5} is substituted as
\EquationHelper{
\prob(\vTheta|\vD) \sim \frac{\exp\left[ -\frac{1}{2} (\vTheta-\hat{\vTheta})^{\top} \hesse (\vTheta-\hat{\vTheta}) \right]}{\left(2\pi \right)^{N\slash 2}\left|\hesse^{-1}\right|^{1\slash 2}},
}{sect2.2.7}
where $\hesse^{-1}$ is the inverse of $\hesse$ and $\left|\bullet\right|$ denotes the determinant of $\bullet$. Equation~\eqref{sect2.2.7} indicates that, in the neighborhood of $\vTheta=\hat{\vTheta}$, $\prob\left(\vTheta|\vD\right)$ can be approximated by a multivariate normal distribution with mean vector $\hat{\vTheta}$ and covariance matrix $\hesse^{-1}$. Let $\Theta_{l}$ ($1\le l \le N$) be a component of interest in $\vTheta$. Integrating Eq.~\eqref{sect2.2.7} over all variables except for $\Theta_{l}$, the marginal distribution with respect to $\Theta_{l}$ is the normal distribution with mean $\hat{\Theta}_{l}$ and variance $(H^{-1})_{l,l}$, which is the $l$-th diagonal element of $\hesse^{-1}$. This means that the uncertainty of $\Theta_{l}$ is given by $(H^{-1})_{l,l}$. When $N\gg1$, it is unrealistic to obtain $\hesse^{-1}$ directly by numerically differentiating $\hesse$, which is generally dense, as this would require $O(CN^2+N^3)$ computations and $O(N^2)$ memory. In practical cases, it is not necessary to evaluate all elements of $\hesse^{-1}$, since the number of elements of interest is usually much smaller than $N$. Therefore, we propose to use a second-order adjoint method~\cite{Wang1998Adjoint, LeDimet2002Second} to efficiently obtain such uncertainties in massive autonomous systems. The following procedure to obtain the uncertainties requires $O(C)$ computations and $O(N)$ memory for each uncertainty. When evaluating the uncertainty of $\Theta_{l}$, we consider a linear equation of $\vc(r)\in\mathbb{R}^{N}$:  
\EquationHelper{
\hesse\vc(r) = \vc(q),
}{sect2.2.8}
where $\vc(q)\in\mathbb{R}^{N}$ is a vector with elements $q_{l}=1$ and $q_{i\ne l}=0$. The solution $\hat{\vc(r)}$ obviously includes $(H^{-1})_{l,l}$ as $\hat{r}_{l} = \sum_{j=1}^{N}(H^{-1})_{l,j} q_{j} = (H^{-1})_{l,l}$. Note that $\hesse$ is a constant matrix that requires complex computations because of its large dimension. We must obtain $\hat{\vc(r)}$ from an initial guess via an iterative technique such as the conjugate gradient method or conjugate residual method. The iterative method needs, in the way of the iteration, to compute each of the Hessian-vector products $\hesse \vgamma$ for a vector $\vgamma$. The second-order adjoint method enables us to compute such Hessian-vector products.

Let $\vxi(t)\in\mathbb{R}^{N}$ and $\vzeta(t)\in\mathbb{R}^{N}$ be perturbations of $\hat{\vtheta}$ and $\hat{\vlambda}$, which respectively correspond to $\vtheta$ and $\vlambda$ when $\vTheta = \hat{\vTheta}$. Their time evolutions are given by  
\EquationHelper{
\frac{\p\vxi}{\p t} =  \left.\frac{\p\vF}{\p\vtheta}\right|_{\vtheta=\hat{\vtheta}} \vxi,
}{sect2.2.9}
\EquationHelper{
\frac{\p\vzeta}{\p t}  + \left.\left(\frac{\p \vF}{\p \vtheta}\right)^{\top} \right|_{\vtheta=\hat{\vtheta}} \vzeta +  \left.\left(\frac{\p^2 \vF}{\p \vtheta^2}\vxi \right)^{\top} \right|_{\vtheta=\hat{\vtheta}} \hat{\vlambda} +\left. \frac{\p^2 \mathcal{J}}{\p\vtheta^2} \right|_{\vtheta=\hat{\vtheta}}\vxi = \vzero.
}{sect2.2.10}
The combination of Eqs.~\eqref{sect2.2.9} and \eqref{sect2.2.10}, which are called the tangent linear model and second-order adjoint model, respectively, gives the Hessian-vector product for an arbitrary vector. Solving Eq.~\eqref{sect2.2.9} forwardly in time for a given vector $\vxi(0)=\vgamma$, we obtain the time series of $\vxi$. Then, solving Eq.~\eqref{sect2.2.10} backwardly with the given $\vzeta(t_{f}) = \vzero$ and the time series $\vxi$, we obtain the objective Hessian-vector product $\vzeta(0) = \hesse \vgamma$. The detailed derivation is given in~\cite{Wang1998Adjoint}.

\section{Twin experiment\label{chap3}}

\subsection{Kobayashi's phase-field model\label{chap3A}}
The accuracy of the proposed method is verified through numerical simulations termed ``twin experiments,'' details of which are given in Section~\ref{chap3B}. We choose a two-dimensional Kobayashi's PF model as the testbed in the twin experiments. Kobayashi's PF model describes the fundamental growth dynamics of two phases, such as in solidification or a phase transformation. The time evolution of one of the phases is described by  
\EquationHelper{
\tau\frac{\p \phi}{\p t} = \epsilon^2 \bigtriangleup \phi + \phi \left(1-\phi\right)\left(\phi + m-\frac{1}{2} \right)\;\text{,}\; - \frac{1}{2}<m<\frac{1}{2}
}{sect3.1.1}
where the PF variable $\phi(\vx,t)$ denotes the existence probability of the relevant phase, e.g., solid or liquid. The parameters $\tau$ and $\epsilon$ non-dimensionalize time and space, respectively, and $m$ characterizes the velocity of the interface between the two phases. We assume that these parameters are time-invariant constants. We know that $\phi(\vx,t)$ should be constrained in $0\le\phi(\vx,t) \le1$, as it describes a probability. This condition is automatically satisfied by setting the initial phase to $0\le\phi(\vx,0)\le1$, since $\phi=0$ and $\phi=1$ are the fixed points of Eq.~\eqref{sect3.1.1}.

Kobayashi's PF model underlies various PF models that describe physical phenomena such as dendrite growth~\cite{kobayashi1993modeling,Shimokawabe2011Peta}, crack propagation~\cite{Aranson2000Continuum,Spatschek2006Phase}, and interface-driven pattern formation~\cite{Komura2007Modelling}. Therefore, Kobayashi's PF model is a good choice for verifying whether the proposed DA method works well, and is a first step towards future applications in more complex PF models.

\subsection{ Synthetic data \label{chap3B}}
Twin experiments are often conducted in the field of DA to verify a newly developed method on the basis of synthetic data. The synthetic data are usually generated using the given simulation model, in which the true parameters and initial state are pre-determined. Verification then proceeds by checking whether the DA method applied to the synthetic data reproduces the true parameters and initial state. In our case, the synthetic dataset is a time series of $\phi$ that is numerically calculated by Kobayashi's PF model with a true initial state and parameter $m$. The synthetic data are then contaminated by observation noise that follows a normal distribution with mean zero and variance $\sigma^2$. The twin experiments are intended to confirm that the proposed method estimates the initial state and parameter with the associated uncertainty.

\begin{figure}[tbp] 
\centering 
\includegraphics[width=0.48\textwidth]{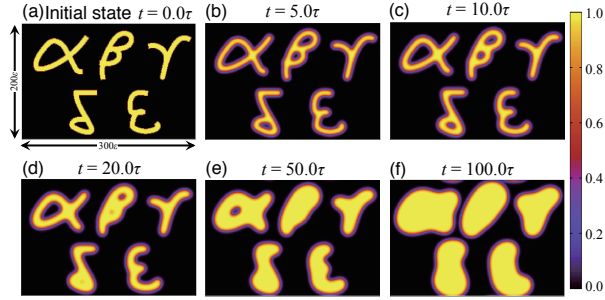} 
\caption{
Time evolution of phase field $\phi$ starting from the initial state shown in (a) for $m=0.1$. 
(b)-(f) show $\phi$ at $t=5.0\tau$, $10.0\tau$, $20.0\tau$, $50.0\tau$, and $100.0\tau$, respectively.
The color indicates the magnitude of $\phi$.
\label{fig:Snaps}}
\end{figure} 

Let $n_{x}$ and $n_{y}$ be the numbers of grid points in the $x$- and $y$-directions, respectively, $M$ be the total number of grid points, i.e., $M=n_x n_y$, and $h$ be the grid spacing. A periodic boundary condition is imposed on the boundary of the computational domain. Letting $\phi_{i}(t)$ be the phase at the $i$-th grid point, Eq.~\eqref{sect3.1.1} can be rewritten as
\EquationHelper{
\tau\frac{\p \phi_{i}}{\p t} = \epsilon^2 \bigtriangleup_{i} \phi_{i} + \phi_{i} \left(1-\phi_{i}\right)\left(\phi_{i} + m-\frac{1}{2} \right), 
}{sect3.2.1}
where $\bigtriangleup_{i}$ denotes a second-order difference operator acting on the four nearest neighbors of the $i$-th grid point $\mathcal{S}_{i}$, i.e., $\bigtriangleup_{i} \phi_{i} = \sum_{j\in\mathcal{S}_{i}}\left( \phi_{j} - \phi_{i}\right)\slash h^2$.

Figure~\ref{fig:Snaps} shows the time evolution of $\phi$ in two-dimensional space, where $n_x=300$, $n_y=200$, $h=\epsilon$, and the time increment in the Euler method is $0.1\tau$. The assumed initial state is shown in Fig.~\ref{fig:Snaps}(a), and the true value for the parameter $m$ is assumed to be $0.1$. Figures~\ref{fig:Snaps}(b)-(f) show snapshots indicating that the interface between the phases $\phi=0$ and $\phi=1$ migrates, expanding the area of $\phi=1$. Motivated by the fact that such snapshots are sometimes obtained as observation data in practical experiments, we use snapshots such as in Figs.~\ref{fig:Snaps}(b)-(f) with added observation noise as the synthetic data for the twin experiments. The synthetic data are given by  
\EquationHelper{
\phi_{i}^{\text{obs}}(t) = \phi_{i}(t) + \omega_{i}(t) \quad(i=1,..,M),
}{sect3.2.2}
where $\omega_{i}(t)$ is normally-distributed observation noise with mean zero and variance $\sigma^2$.

When DA is applied to Kobayashi's PF model, the time evolution equation with respect to $m$ is needed to construct a system model within the state-space model (Section~\ref{chap2A}). The time evolution equation can be written as  
\EquationHelper{
\tau \frac{\p b}{\p t} = 0,
}{sect3.2.3}
where $b=m+1\slash 2$, which denotes the normalization of $m$, i.e., $0<b<1$.

\subsection{Cost function\label{chap3C}}
We consider the synthetic observation data to be the snapshots of $\phi$ obtained from $t=T_{\text{min}}$ to $t=T_{\text{max}}$ with time interval $\Delta T$. Let $\mathcal{T}$ be the set of observation times and $n$ be the number of observations. Combining Eqs.~\eqref{sect2.1.10} and \eqref{sect3.2.2}, $J$ can be rewritten as  
\EquationHelper{
J = \frac{nM}{2}\log(2\pi\sigma^2)+ \frac{1}{2\sigma^2}\sum_{t_{s}\in \mathcal{T}}\sum_{i=1}^{M}\left( \phi^{\text{obs}}_{i}(t_s) - \phi_{i}(t_s) \right)^2.
}{sect3.3.1}
The values of $\phi_{i}(0)$ and $b(0)$ that minimize Eq.~\eqref{sect3.3.1} also minimize  
\EquationHelper{
J^{\prime} = \int_{0}^{T}dt\;   \mathcal{J}^{\prime},
}{sect3.3.3}
where 
\EquationHelper{
\mathcal{J}^{\prime}= \frac{1}{2}\sum_{t_{s}\in \mathcal{T}}\delta(t-t_{s}) \sum_{i=1}^{M}\left( \phi^{\text{obs}}_{i}(t) - \phi_{i}(t) \right)^2,
}{sect3.3.2}
since $\sigma$ is independent of $\phi_{i}(0)$ and $b(0)$. When the optimum $\hat{\phi}_{i}(0)$ for $\phi_{i}(0)$ and $\hat{b}(0)$ for $b(0)$ are obtained by minimizing $J^{\prime}$, the optimum $\sigma$ can be obtained as 
\EquationHelper{
\hat{\sigma} = \sqrt{ \frac{1}{nM} \sum_{t_{s}\in \mathcal{T}} \sum_{i=1}^{M}\left( \phi^{\text{obs}}_{i}(t_s) - \hat{\phi}_{i}(t_s)\right)^2 },
}{sect3.3.4}
where $\hat{\phi}_{i}(t)$ denotes $\phi_{i}(t)$ simulated using $\hat{\phi}_{i}(0)$ and $\hat{b}(0)$.

\subsection{Procedures\label{chap3D}}
Prior to applying the proposed method to the PF model, the constraint for the initial state $0\le\phi_{i}(0)\le 1$ is to be changed to $0<\phi_{i}(0)<1$ to satisfy the domain of the variable transformation Eq.~\eqref{variabletrainsformation}. The state variables $\vtheta(t)\in \mathbb{R}^{M+1}$ and $\vTheta\in\mathbb{R}^{M+1}$ can be defined as 
\EquationHelper{
\begin{aligned}
\vtheta & = \left( \phi_{1},\cdots,\phi_{M}, b \right)^{\top} \\
\vTheta & = \left( \phi_{1}(0),\cdots,\phi_{M}(0), b(0) \right)^{\top}.
\end{aligned}
}{aa}
The constraint for $\vTheta$ becomes $0<\Theta_{i}<1$ ($i=1,\cdots,M+1$). The system models of Eqs.~\eqref{sect3.2.1} and \eqref{sect3.2.3} are rewritten in terms of $\vtheta$ as
\EquationHelper{
\tau\frac{\p \theta_{i}}{\p t} = \left\{
\begin{array}{ll}
 \epsilon^2 \bigtriangleup_{i} \theta_{i} + \theta_{i} \left(1-\theta_{i}\right)\left(\theta_{i} + \theta_{M+1} -1 \right) & \text{for} \;\;i=1,\cdots,M, \\
 0 & \text{otherwise}.
\end{array}\right.
}{sect3.4.1}
Substituting the right-hand side of this equation for $\vF$ in Eq.~\eqref{sect2.2.3}, replacing $\mathcal{J}$ (Eq.~\eqref{sect2.2.3}) with $\mathcal{J}^{\prime}$ (Eq.~\eqref{sect3.3.2}), and replacing $J$ (Eq.~\eqref{sect2.2.4}) with $J^{\prime}$ (Eq.~\eqref{sect3.3.3}), the adjoint method described by Eqs.~\eqref{sect2.2.3}-\eqref{sect2.2.4-1} is rewritten as
\EquationHelper{
-\tau\frac{\p \lambda_{i} }{\p t} = \left\{
\begin{aligned}
& \epsilon^2 \bigtriangleup_{i} \lambda_{i}  + \left\{-3\theta_{i}^2 + \left(4-2\theta_{M+1}\right)\theta_{i} + \theta_{M+1}-1\right\} \lambda_{i} +  \frac{\p \mathcal{J}^{\prime} }{\p\theta_{i}}
 \quad\text{for} \;\;i=1,\cdots,M, \\
& \sum_{j=1}^{M}\theta_{j}\left(1-\theta_{j}\right)\lambda_{j} \qquad\qquad\text{otherwise},
\end{aligned}\right.
}{sect3.4.2}
\EquationHelper{
\vlambda(0) = \frac{\p J^{\prime}}{\p \vTheta},
}{sect3.4.2-1}
\EquationHelper{
\vlambda(t_{f}) = \vzero.
}{sect3.4.2-2}

We adopt the LBFGS technique~\cite{nocedal1980updating} as the gradient method for optimizing $\vTheta$. Starting from an initial guess, the LBFGS method updates $\vTheta$ by satisfying $0<\Theta_{i}<1$ for all $i$ owing to the variable transformation mentioned in Section~\ref{chap2B}. To tune the LBFGS method, we set the tolerance to $10^{-8}$ and determine the step length by Armijo's rule~\cite{Armijo1966Minimization}. Once the optimum $\hat{\vTheta}$ has been obtained, the optimum standard deviation $\hat{\sigma}$ can be estimated by Eq.~\eqref{sect3.3.4} and the optimum $\hat{m}$ for $m$ is given by $\hat{\Theta}_{M+1} - 1\slash2$ or $\hat{b}(0)-1\slash2$.

One of the most remarkable features of the proposed method is its evaluation of the uncertainties. These uncertainties can provide important information that is beneficial to updating the experimental design. In accordance with the procedure mentioned in Section~\ref{chap2D}, we consider a linear equation ${\vH}^{\prime} \vr = \vq$, where ${\vH}^{\prime}$ = $\left.\p^2 J^{\prime} \slash \p \vTheta^2\right|_{\vTheta=\hat{\vTheta}}$ is a Hessian matrix, $\vr \in \mathbb{R}^{M+1}$ is a vector to be determined, and $\vq \in \mathbb{R}^{M+1}$ is a vector containing the elements $q_{M+1}=1$ and $q_{i\ne M+1}=0$. The uncertainty $\delta \hat{m}$ can be computed from the solution $\hat{\vr}$ as
\EquationHelper{
\delta \hat{m} = \hat{\sigma}\sqrt{\hat{r}_{M+1}}.
}{delm}
The conjugate residual method, in which the tolerance is set to $10^{-8}$, is adopted to solve the linear equation. The second-order adjoint method computes each of the Hessian-vector products ${\vH}^{\prime}\vgamma$ that appears in the optimization process of the conjugate residual method. Substituting the right-hand side of Eq.~\eqref{sect3.4.1} for $\vF$ in Eq.~\eqref{sect2.2.9}, the tangent linear model can be rewritten as 
\EquationHelper{
\tau\frac{\p \xi_{i} }{\p t} = \left\{
\begin{aligned}
& \epsilon^2 \bigtriangleup_{i} \xi_{i} + \hat{\theta}_{i}\left(1-\hat{\theta}_{i}\right)\xi_{M+1} + \left\{-3 \hat{\theta}_{i}^2 + \left(4-2\hat{\theta}_{M+1}\right)\hat{\theta}_{i} + \hat{\theta}_{M+1}-1\right\} \xi_{i} \\
& \qquad\qquad\qquad\quad\text{for} \;\;i=1,\cdots,M, \\
& 0  \;\qquad\qquad\qquad\text{otherwise},
\end{aligned}\right.
}{sect3.4.3}
with the initial condition $\vxi(0)=\vgamma$, where $\hat{\vtheta}$ denotes the state vector corresponding to $\hat{\vTheta}$. Substituting the right-hand side of Eq.~\eqref{sect3.4.1} for $\vF$ in Eq.~\eqref{sect2.2.10} and replacing $\mathcal{J}$ (Eq.~\eqref{sect2.2.10}) with $\mathcal{J}^{\prime}$ (Eq.~\eqref{sect3.3.2}), the second-order adjoint model in Eq.~\eqref{sect2.2.10} becomes
\EquationHelper{
- \tau\frac{\p \zeta_{i} }{\p t} = \left\{
\begin{aligned}
 \epsilon^2 & \bigtriangleup_{i} \zeta_{i} \\
& + \left\{-3 \hat{\theta}_{i}^2 + \left(4-2\hat{\theta}_{M+1}\right)\hat{\theta}_{i} + \hat{\theta}_{M+1}-1\right\} \zeta_{i} \\
& - \left(6\hat{\theta}_{i} + 2\hat{\theta}_{M+1} -4 \right) \hat{\lambda}_{i}\xi_{i} - \left( 2\hat{\theta}_{i} - 1\right) \hat{\lambda}_{i}\xi_{M+1}  \\ 
& + \left. \sum_{j=1}^{M} \frac{\p^{2} \mathcal{J}^{\prime} }{\p\theta_{i}\p\theta_{j}} \right|_{\vtheta=\hat{\vtheta}}\xi_{j} 
 \qquad\qquad\qquad\qquad\text{for} \;\;i=1,\cdots,M, \\
&  \sum_{j=1}^{M}\left[ \hat{\theta}_{j}\left(1-\hat{\theta}_{j}\right)\zeta_{j} - \left( 2\hat{\theta}_{j} - 1\right) \hat{\lambda}_{j}\xi_{j} \right] 
  \quad\text{otherwise},
\end{aligned}\right.
}{sect3.4.4}
where $\hat{\vlambda}$ is the perturbation of $\hat{\vtheta}$. Solving Eq.~\eqref{sect3.4.4} with the condition $\vzeta(t_{f})=\vzero$, we obtain the objective Hessian-vector product as $\vzeta(0)= {\vH}^{\prime}\vgamma$.

\section{Results and discussion\label{chap4}}
The proposed method is verified through three twin experiments: ({\rI}) estimation of the parameter $m$ conditional on the true initial state $\phi_{i}^{\text{true}}(0)$ (Section~\ref{chap4A}), ({\rII}) simultaneous estimation of the parameter $m$ and the initial state $\phi_{i}(0)$ (Section~\ref{chap4B}), and ({\rIII}) estimation of the initial state $\phi_{i}(0)$ conditional on the true parameter $m^{\text{true}}$ (Section~\ref{chap4C}). Twin experiment {\rI} investigates how the estimation depends on observation data, twin experiment {\rII} verifies whether the proposed method outputs correct estimations, even for massive simulation models, and twin experiment {\rIII} validates the unknown phenomena that appear in the results of experiment {\rII}.

\subsection{Twin experiment {\rI}: Parameter estimation\label{chap4A}}
Twin experiment I investigates the influences of three parameters related to the observation data: (i) the length of the observation time $T_{\text{max}}$, (ii) the time interval of the observations $\Delta T$, and (iii) the standard deviation of the observation noise $\sigma$. Table~\ref{tableparam} summarizes the parameter values used in the experiments. The first observation is assumed to occur at $T_{\text{min}}=0.1\tau$, and the true value of $m$ is assumed to be $m^{\text{true}}=0.1$. The initial guess used in the LBFGS method is $m=-0.1$. In this experiment, the true phase field at $t=0$, i.e., $\phi^{\text{true}}_{i}(0)$ (see Fig.~\ref{fig:Snaps}(a)) is given as the initial state. The results reported here are the average values of $\hat{m}$ and $\delta  \hat{m}$ over twenty trials with different random seeds.

\begin{table}[tb]
\centering
  \begin{tabular}{|c|c|c|c|}
 \hline
& $T_{\text{max}}$ & $\Delta T$ & $\sigma$ \\ \hline
Test \rI-(i) & $0.2\tau - 102.4\tau$ & $0.1\tau$ & $0.01$ \\ \hline
Test \rI-(ii) & $102.5\tau$ & $0.1\tau - 51.2\tau$ & $0.01$ \\ \hline
Test \rI-(iii) & $102.4\tau$ & $0.1\tau$ & $10^{-5}- 1.0$ \\ \hline
  \end{tabular}
\caption{Length of the observation time $T_{\text{max}}$, time interval of observations $\Delta T$, and standard deviation of observation noise $\sigma$ used in twin experiment {\rI}, where $\tau$ is the unit of time in the simulation. Test {\rI}-(i), (ii) and (iii) investigate how the estimation depends on $T_{\text{max}}$, $\Delta T$ and $\sigma$, respectively.}\label{tableparam}
\end{table}

\begin{figure}[tbp] 
\centering 
\includegraphics[width=0.48\textwidth]{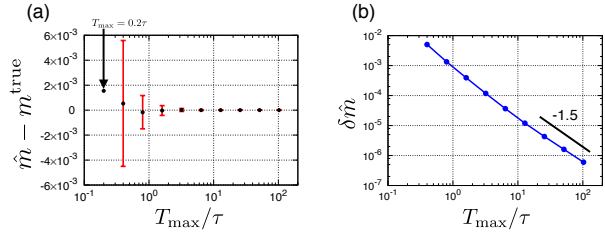} 
\caption{ Results of twin experiment {\rI}-(i). The length of each error bar for the optimum parameter $\hat{m}$ in (a) corresponds to the estimated uncertainty $\delta\hat{m}$ in (b).
The uncertainty cannot be determined when $T_{\text{max}}=0.2\tau$.
The black solid line in (b) indicates a power function of order $-1.5$.
\label{fig:Tmax}}
\end{figure} 

Figure~\ref{fig:Tmax} shows the results of Test {\rI}-(i). Figure~\ref{fig:Tmax} (a) indicates that the parameter estimation is successful, because the true parameter is included in the range $\hat{m} - \delta \hat{m} < m^{\text{true}}<\hat{m} + \delta \hat{m}$. The estimation of the uncertainty $\delta\hat{m}$ fails when $T_{\text{max}}$ is at its minimum, i.e., $T_{\text{max}}= 0.2\tau$, because insufficient data cause $\hat{r}_{M+1}$ to become negative. Figure~\ref{fig:Tmax}(b) indicates that $\delta \hat{m}$ is proportional to $T_{\text{max}}^{-1.5}$ in this range. The reason that the decrease is more rapid than the law of large numbers would suggest is related to the nonlinearity of  Kobayashi's PF model. A theoretical evaluation actually indicates that $\delta \hat{m}$ is proportional to $T_{\text{max}}^{-2.5}$ when $T_{\text{max}}\ll \tau$, and will converge with a constant value as $T_{\text{max}}$ increases. This is because no additional information is included in the observation data after $\phi(\vx,t)$ becomes almost uniform across the entire computational domain.
\begin{figure}[tbp] 
\centering 
\includegraphics[width=0.48\textwidth]{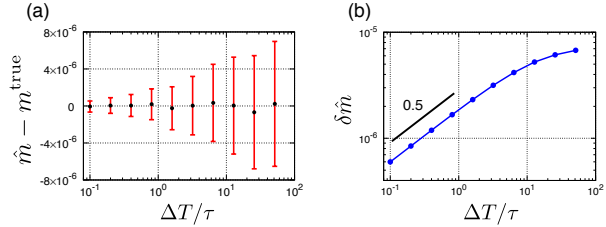} 
\caption{
Results of twin experiment {\rI}-(ii). The length of each error bar for optimum parameter $\hat{m}$ in (a) corresponds to the estimated uncertainty $\delta\hat{m}$ in (b). The black solid line in (b) indicates a power function of order of $0.5$.
\label{fig:DelT}}
\end{figure} 
Figure~\ref{fig:DelT} shows the results of Test {\rI}-(ii). Figure~\ref{fig:DelT}(a) indicates that the proposed method successfully reproduces the true parameter, and Fig.~\ref{fig:DelT}(b) shows that $\delta \hat{m}$ is proportional to $\Delta T^{0.5}$ when $\Delta T<\tau$, which seems to follow the law of large numbers.
\begin{figure}[tbp] 
\centering 
\includegraphics[width=0.48\textwidth]{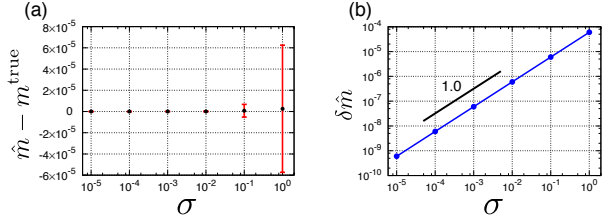} 
\caption{
Results of twin experiment {\rI}-(iii). The length of each error bar for optimum parameter $\hat{m}$ in (a) corresponds to the estimated uncertainty $\delta\hat{m}$ in (b). The black solid line in (b) indicates a linear function.
\label{fig:Sigma}}
\end{figure} 
Figure~\ref{fig:Sigma} shows the results of Test {\rI}-(iii). Figure~\ref{fig:Sigma}(a) indicates that the parameter estimation is again successful, and Fig.~\ref{fig:Sigma}(b) shows that $\delta \hat{m}$ is proportional to $\sigma$. In summary, the results of Test {\rI} demonstrate that the proposed method is capable of estimating the true parameter and the associated uncertainty.

\subsection{Twin experiment {\rII}: Simultaneous estimation\label{chap4B}}
Twin experiment {\rII} investigates the influence of the observation noise in two cases: (i) when the noise has a small standard deviation ($\sigma=10^{-4}$) and (ii) when the noise has a large standard deviation ($\sigma=0.3$). The true parameter is assumed to be $m^{\text{true}}=0.1$, and the true initial state $\phi_{i}^{\text{true}}(0)$ is assumed to be the phase field shown in Fig.~\ref{fig:Snaps}(a). The other observational conditions are $T_{\text{min}}= 5.0\tau$, $T_{\text{max}}=30.0\tau$, $\Delta T=0.1\tau$, and the initial guesses are $\phi_{i}(0)=0.2$ and $m=-0.2$.

\begin{figure}[tbp] 
\centering 
\includegraphics[width=0.48\textwidth]{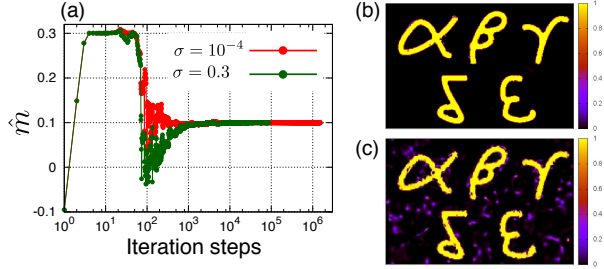} 
\caption{
Results of twin experiment {\rII}. (a) How the LBFGS method updates $m$ when the observation data noise has a small (red) and large (green) standard deviation. (b) The optimum initial state of the phase field $\phi_{i}(0)$ in the case of small noise, and (c) that in the case of large noise.
\label{fig:simul}}
\end{figure} 
Figure~\ref{fig:simul} shows the results of Test {\rII}. Figure~\ref{fig:simul}(a) indicates how each iteration of the LBFGS method updates the estimation of $m$. It is clear that each estimation converges with $m^{\text{true}}$. Figures~\ref{fig:simul}(b) and (c) indicate the estimated initial states $\hat{\phi}_{i}(0)$ in Test {\rII}-(i) and (ii), respectively. These results appear to be almost consistent with the true initial states, although ``spot-like'' pattern appears in Fig.~\ref{fig:simul}(c). This spot-like pattern would be conspicuous if the observation noise was large or if the time of the first observation $T_{\text{min}}$ was far from $t=0$. Additionally, the spot-like pattern does not disappear under lower tolerance levels.

\subsection{Twin experiment {\rIII}: Estimation of initial state\label{chap4C}}
Twin experiment {\rIII} confirms whether the estimation of $m$ affects the generation of the spot-like pattern found in twin experiment {\rII}-(ii). Therefore, twin experiment {\rIII} is set up to estimate only the initial state $\phi_{i}(0)$ with a fixed parameter $m=0.1$. The true initial state $\phi_{i}^{\text{true}}(0)$ is assumed to be the phase field shown in Fig.~\ref{fig:Snaps}(a). The other observational conditions are $T_{\text{min}}= 8.0\tau$, $T_{\text{max}}=30.0\tau$, $\Delta T=0.1\tau$ and $\sigma=0.3$, and the initial guess is $\phi_{i}(0)=0.2$.

\begin{figure}[tbp] 
\centering 
\includegraphics[width=0.48\textwidth]{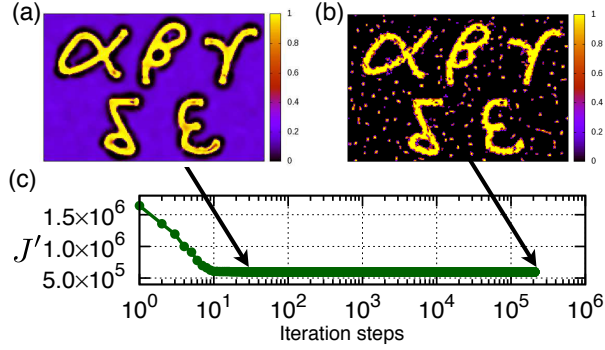} 
\caption{ Results of twin experiment {\rIII}.
Estimated initial states of the phase field $\phi_{i}(0)$ (a) after the $31$st step and (b) after the final step in the iteration of the LBFGS method.
(c) Improvement in the cost function $J^{\prime}$.
\label{fig:spots}}
\end{figure} 
Figures~\ref{fig:spots}(a) and (b) show the estimated initial states after the $31$st and after the final iterations, respectively, and Fig.~\ref{fig:spots}(c) shows how the cost function $J^{\prime}$ varies with the iteration. A spot-like pattern again appears in the estimated initial state (Fig.~\ref{fig:spots}(b)) as the number of iterations increases. Note that the cost function $J^{\prime}$ is almost the same after the $31$st step and after the final step, although Figs.~\ref{fig:spots}(a) and (b) are much different.

This is caused by a feature inherent in the two-dimensional Kobayashi's PF model. When a spot of radius $R_{0}$ evolves with time based on the PF model, whether it grows or decays depends on the relation between $m$ and $R_{0}$.
\begin{figure}[tbp] 
\centering 
\includegraphics[width=0.48\textwidth]{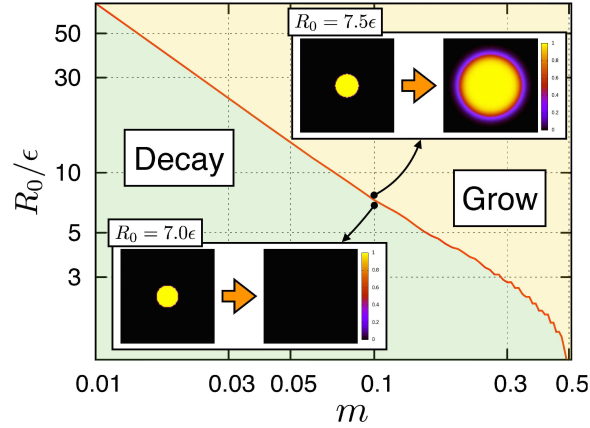} 
\caption{Phase diagram for an axisymmetric two-dimensional Kobayashi's PF model. The red line indicates the critical radius as a function of the parameter $m$. The region above or below the critical line corresponds to a spot growing or decaying with time, respectively.\label{fig:phasediagram}}
\end{figure} 
Figure~\ref{fig:phasediagram} shows the phase diagram obtained by the two-dimensional Kobayashi's PF model under the assumption of the axial symmetry. The destiny of a given spot depends on whether the radius is above or below the critical line, which means the critical radius is approximately inversely proportional to $m$~\cite{Castro2003Phase}. The radius of each spot in Fig.~\ref{fig:spots}(b) is actually smaller than the critical radius, which is approximately $7.3\epsilon$ in the case of $m=0.1$.

\begin{figure}[tbp] 
\centering 
\includegraphics[width=0.48\textwidth]{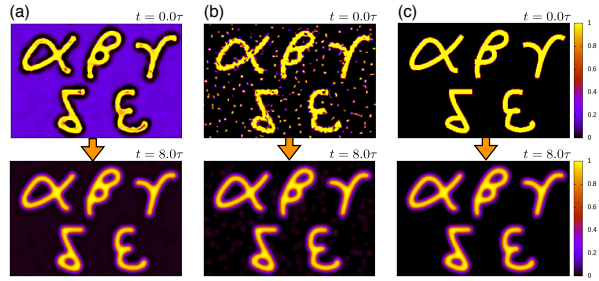} 
\caption{Time evolutions of the phase fields starting from different initial states $\phi_{i}(0)$:
(a) the estimated initial state obtained after the $31$st step (Fig.~\ref{fig:spots}(a)), (b) that after the final step (Fig.~\ref{fig:spots}(b)), and (c) the true initial state (Fig.~\ref{fig:Snaps}(a)).
\label{fig:spotsfwd}}
\end{figure} 
Time evolutions starting from three different initial states shown in Fig.~\ref{fig:spots}. These results indicate that the phase fields at the time of the first observation, i.e., $t=8.0\tau$, are completely coincident (Fig.~\ref{fig:spotsfwd}).

\section{Conclusions\label{chap5}}
This paper has described an adjoint-based DA method for massive autonomous models that not only determines the optimum estimates but also gives their uncertainties within a practical computation time and reasonable resource requirements. The uncertainties can be obtained as several diagonal components of the inverse Hessian matrix, which is the covariance matrix of the normal distribution that approximates the posterior PDF in the neighborhood of the optimum estimates. This proposed approach provides a new methodology for evaluating uncertainties using a second-order adjoint method that obtains Hessian-vector products in the process of the computation. Twin experiments using a two-dimensional Kobayashi PF model demonstrated the validity of the proposed method.

The uncertainties associated with physical quantities of interest depend on the quality and amount of data. Thus, conducting twin experiments prior to practical experiments allows us to determine how many observations are required to obtain the physical quantities of interest to the desired accuracy. Such feedback to practical experiments is already possible in systems with only a few degrees of freedom, but the proposed method makes this possible for massive simulation models.

The proposed method is not only applicable to PF models, but to various models described by autonomous systems, e.g., shallow water equations, Navier equations for elastic materials, and Boltzmann equations. The proposed method is of great utility for evaluating the uncertainties of model parameters through DA, which is important in various fields of science, even when using massive models.

\section*{Acknowledgments}
This study is supported by the Cross-ministerial Strategic Innovation Promotion Program (SIP). The authors are grateful to Prof. Munekazu Ohno, Prof. Peter XK Song and Dr. Jonggyu Baek for useful discussions.

\end{document}